\documentclass[twocolumn,twocolappendix]{aastex631}
\usepackage{showyourwork}

\usepackage{lipsum} 

\usepackage{tikz}
\usetikzlibrary{positioning}
\usetikzlibrary{shapes}
\usetikzlibrary{fit}

\newcommand{\comment}[1]{}

\newcommand{\px}{p^{}_{X}}
\newcommand{\pu}{p^{}_{U}}
\newcommand{\R}{\mathbb{R}}

\defcitealias{schmidt2020}{Schmidt \& Malz et al.}
\newcommand{\citePZp}{(\citetalias{schmidt2020} \citeyear{schmidt2020})\xspace}
\newcommand{\citePZt}{\citetalias{schmidt2020} (\citeyear{schmidt2020})\xspace}
\newcommand{\citePZa}{\citetalias{schmidt2020} \citeyear{schmidt2020}\xspace}

\newcommand{\dirac}{DIRAC Institute and the Department of Astronomy, University of Washington, Seattle, WA 98195, USA}

\shorttitle{Galaxy Catalogues with Normalizing Flows}
\shortauthors{Crenshaw et al.}

\begin{document}

\title{Probabilistic Forward Modeling of Galaxy Catalogs with Normalizing Flows}

\correspondingauthor{John~Franklin~Crenshaw}
\email{jfc20@uw.edu}

\author[0000-0002-2495-3514]{John~Franklin~Crenshaw}
\affiliation{\dirac}
\affiliation{Department of Physics, University of Washington, Seattle, WA 98195, USA}

\author[0000-0002-6825-5283]{J.~Bryce~Kalmbach}
\affiliation{\dirac}

\author[0000-0003-4906-8447]{Alexander~Gagliano}
\affiliation{Department of Astronomy, University of Illinois at Urbana-Champaign, 1002 W. Green St., IL 61801, USA}
\affiliation{National Center for Supercomputing Applications, Urbana, IL, 61801, USA}
\affiliation{Center for AstroPhysical Surveys, Urbana, IL, 61801, USA}
\affiliation{National Science Foundation Graduate Research Fellow}

\author[0000-0001-8043-5378]{Ziang Yan}
\affiliation{Ruhr University Bochum, Faculty of Physics and Astronomy, Astronomical Institute (AIRUB), German Centre for Cosmological Lensing, 44780 Bochum, Germany}

\author[0000-0001-5576-8189]{Andrew~J.~Connolly}
\affiliation{\dirac}

\author[0000-0002-8676-1622]{Alex~I.~Malz}
\affiliation{McWilliams Center for Cosmology, Department of Physics, Carnegie Mellon University}

\author[0000-0002-5091-0470]{Samuel~J.~Schmidt}
\affiliation{Department of Physics and Astronomy, University of California, One Shields Avenue, Davis, CA 95616, USA}

\author{The LSST Dark Energy Science Collaboration}

\begin{abstract}
    Evaluating the accuracy and calibration of the redshift posteriors produced by photometric redshift (photo-z) estimators is vital for enabling precision cosmology and extragalactic astrophysics with modern wide-field photometric surveys.
    Evaluating photo-z posteriors on a per-galaxy basis is difficult, however, as real galaxies have a true redshift but not a true redshift posterior.
    We introduce PZFlow, a Python package for the probabilistic forward modeling of galaxy catalogs with normalizing flows.
    For catalogs simulated with PZFlow, there is a natural notion of ``true'' redshift posteriors that can be used for photo-z validation.
    We use PZFlow to simulate a photometric galaxy catalog where each galaxy has a redshift, noisy photometry, shape information, and a true redshift posterior.
    We also demonstrate the use of an ensemble of normalizing flows for photo-z estimation.
    We discuss how PZFlow will be used to validate the photo-z estimation pipeline of the Dark Energy Science Collaboration (DESC), and the wider applicability of PZFlow for statistical modeling of any tabular data.
\end{abstract}

\section{Introduction}
\label{sec:intro}

\begin{figure*}[t]
    \script{intro/plot_two_moons.py}
    \begin{centering}
        \includegraphics{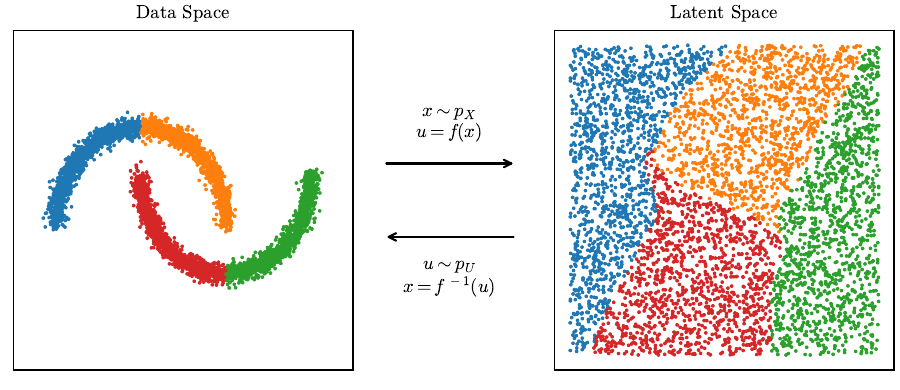}
        \caption{
            A normalizing flow demonstrated on the two moons data set from scikit-learn.
            The two moons data on the left is mapped onto a two dimensional uniform distribution by the bijection $f$.
            The data are colored by quadrant to visualize their image in the latent space.
            You can sample the data distribution by sampling from the uniform distribution, and using $f^{-1}$ to map the samples back to the data space.
        }
        \label{fig:two-moons}
    \end{centering}
\end{figure*}

Photometric redshift (photo-z) estimation is necessary for the study of cosmology and galaxy evolution with the huge galaxy catalogs generated by modern wide-field photometric surveys like the Vera C. Rubin Observatory's Legacy Survey of Space and Time (LSST; \citealt{ivezic2019}).
The colors of galaxies in these surveys are used to estimate posterior distributions over the possible redshift of each galaxy.
These posteriors are often multimodal, reflecting degeneracies in redshift and galaxy spectral type.
Guaranteeing the accuracy and calibration of these posteriors, and understanding their biases, is vital for enabling precision cosmology and studies of galaxy evolution \citep{newman2022,descSRD}.

There have been many studies evaluating the performance of photo-z estimation by comparing estimates to known true redshifts (e.g., \citealt{hildebrandt2010}, \citealt{sanchez2014}, \citealt{graham2018}), however to meet the needs of modern surveys, we must evaluate the accuracy of the full redshift posteriors generated for these galaxies.
In other words, not only must we verify that photo-z estimators assign a high posterior probability to the true redshift, but also that the posteriors account for all the degeneracies inherent in photo-z estimation, and that the distributions are neither too narrow nor too broad.
This kind of validation is difficult, however, because individual galaxies do not have a ``true'' redshift posterior to which you can compare the estimated posterior.
\citePZt studied whether these questions could be answered using ensemble-level comparisons of photo-z posteriors and true redshifts, but found that these metrics could be fooled by a pathological photo-z estimator that ignored galaxy colors altogether.
This raised the need for new methodologies that enable the evaluation of photo-z posterior estimates at the per-galaxy level.

One way to enable posterior validation at the per-galaxy level is to simulate catalogs by sampling galaxies from a probabilistic model for which you have direct access to the underlying probability distribution.
Then, for each galaxy, you can calculate the posterior over redshift directly from the same model that generated the galaxy.
This provides a ``true'' redshift posterior for each simulated galaxy to which you can compare the posteriors from photo-z estimators, enabling the validation of individual photo-z posterior estimates.

Normalizing flows are a generative model of this variety.
Normalizing flows model complex, high-dimensional probability distributions using deep neural networks that learn an invertible mapping between the complicated data distribution and a more simple distribution, known as the latent distribution.
This allows you to sample from and calculate probabilities with respect to the latent distribution, and use the normalizing flow to translate these values back to the space of the original data distribution.
A common choice for the latent distribution is the Normal distribution, hence the name \emph{normalizing} flow.

Unlike other deep generative models like Generative Adversarial Networks (GANs; \citealt{goodfellow2014}) and Variational Autoencoders (VAEs; \citealt{kingma2014}), normalizing flows provide a deterministic mapping between the data and latent spaces.
It is this feature of normalizing flows that enables us to provide an exact answer to the question ``under the generative model, what is the posterior distribution for redshift given the simulated photometry''.
This allows us to define a ``true'' redshift posterior for each galaxy in our simulated catalog, making normalizing flows a very powerful tool for validating photo-z inference.

In this paper, we introduce PZFlow, a normalizing flow package for Python that is designed to facilitate forward modeling galaxy catalogs with true posteriors for galaxy properties.
In addition, we demonstrate PZFlow as a photo-z estimator.
With relatively little tuning required by the user, PZFlow can provide a generative model for any tabular data, including continuous and discrete variables, and variables with Euclidean or periodic topology (e.g. the celestial sphere).
While calculating posteriors, PZFlow can convolve error distributions and marginalize over missing values.

In Section~\ref{sec:nf} we provide the background on normalizing flows required to understand PZFlow, which we describe in Section~\ref{sec:pzflow}.
In Section~\ref{sec:galaxy-catalog}, we demonstrate using PZFlow to simulate a galaxy catalog where each object has a redshift, photometry in the six LSST bands, a true photo-z posterior, a size, and an ellipticity.
In Section~\ref{sec:photo-z}, we demonstrate using PZFlow as a density estimator by estimating photo-z's for our simulated catalog.
We conclude in Section~\ref{sec:conclusion}.

\section{Normalizing Flows}
\label{sec:nf}

Normalizing flows model complex, high-dimensional probability distributions by learning a mapping from the data distribution to a tractable latent distribution\footnote{Some of the machine learning literature defines the mapping in the opposite direction.}.
Often the latent distribution is a standard Normal distribution, and so the mapping ``normalizes'' the data, hence the name ``normalizing flow''.
This mapping allows us to sample and evaluate densities using the latent distribution, rather than the unknown data distribution.

Assume we have a differentiable function $f$ that maps samples $x$ from the data distribution $\px$ onto samples $u$ from the latent distribution $\pu$.
Using the change of variables formula, we can evaluate the probability density of the data:
\begin{align}
    \px(x) = \pu(u=f(x)) \, |\det \nabla f(x)|,
    \label{eq:px}
\end{align}
where $\nabla f(x)$ is the Jacobian of $f$ evaluated at $x$.
In words, computing the density $\px(x)$ is accomplished by mapping $x$ to the latent distribution, calculating its density there, and multiplying by the associated Jacobian determinant, which accounts for how the function $f$ distorts volume elements of the space.

If we further assume that $f$ is invertible, we can sample from the data distribution by applying $f^{-1}$ to samples from the latent distribution\footnote{Here, $\sim$ means ``is drawn from.''}:
\begin{align}
    x = f^{-1}(u) \quad \text{where} \quad u \sim \pu.
\end{align}

Figure~\ref{fig:two-moons} shows an example of a normalizing flow that transforms the scikit-learn \citep{sklearn} two moons distribution into a uniform distribution.
The data points are colored by quadrant to visualize their image under $f$.

The following sections discuss how to build a normalizing flow to model data with various features.
Section~\ref{sec:bijections} discusses the bijection $f$ and introduces the building blocks from which our bijections will be built;
Section~\ref{sec:latent} discusses how to choose an appropriate latent distribution for your data;
Section~\ref{sec:conditional} describes how to build a flow that models a conditional distribution;
Section~\ref{sec:periodic} explains how to model data with periodic topology;
finally Section~\ref{sec:discrete} explains how to model data with discrete variables.

\subsection{Designing a bijection}
\label{sec:bijections}

A bijection is an invertible map between two sets.
In a normalizing flow, the bijection maps the data distribution onto the latent distribution for probability calculation, and the inverse of the bijection maps samples from the latent distribution back to the data distribution.
The bijection of a normalizing flow must be powerful enough to model complex relationships in data, while remaining invertible and simultaneously possessing an efficiently computable Jacobian determinant.
This latter constraint is the primary difficulty in designing a normalizing flow.
The most popular strategy for achieving these requirements is to exploit the fact that a composition of bijections is also bijective.
By chaining together multiple less-expressive bijections whose Jacobians are efficiently computable, a composite bijections can be constructed that meets our requirements:
\begin{align}
    f &= \dots \circ f_3 \circ f_2 \circ f_1.
\end{align}
The overall Jacobian determinant can be efficiently calculated using the chain rule.

There is an extensive literature on constructing these sub-bijections (see \citealt{kobyzev2020} for a review).
Some bijections are specialized to be particularly efficient at either density estimation or sampling, but for many science cases, we wish to do both.
For this reason, we will focus on Rational-Quadratic Rolling Spline Couplings (RQ-RSCs), bijections which achieve state-of-the-art performance, while being efficient with both tasks \citep{durkan2019}.

\subsubsection{Rational-Quadratic Rolling Spline Couplings}
\label{sec:rq-rsc}

\begin{figure}
    \centering
    \begin{tikzpicture}[thick, outer sep=0]
        \node [draw, rectangle, minimum height=2cm, minimum width=1.28cm, fill=blue!20] (1) {$x^{}_{1:d}$};
        \node [draw, circle, minimum size=8mm, fill=gray!20, right = 2cm of 1] (2) {$=$};
        \node [draw, rectangle, minimum height=2cm, minimum width=1.28cm, fill=blue!20, right = 1.5cm of 2] (3) {$y^{}_{1:d}$};
        \draw[->] (1) -- (2);
        \draw[->] (2) -- (3);
        \node [draw, rectangle, minimum height=2cm, minimum width=1.28cm, fill=red!20, below = 0 of 1] (4) {$x^{}_{d+1:D}$};
        \node [draw, circle, minimum size=8mm, fill=gray!20, right = 2cm of 4] (5) {$g$};
        \node [draw, rectangle, minimum height=2cm, minimum width=1.28cm, fill=red!20, below = 0 of 3] (6) {$y^{}_{d+1:D}$};
        \draw[->] (4) -- (5);
        \draw[->] (5) -- (6);
        \draw[->] (1) -- (5) node [midway, draw, circle, fill=white, minimum size=8mm, fill=gray!20] (7) {$m$};
    \end{tikzpicture}
    \caption{
        Diagram of a coupling layer.
        The first partition, $x^{}_{1:d}$, is passed through the layer unchanged.
        The second partition, $x^{}_{d+1:D}$, is transformed by the coupling law $g$, which is parameterized by the coupling function $m$ applied to the first partition.
    }
    \label{fig:coupling}
\end{figure}
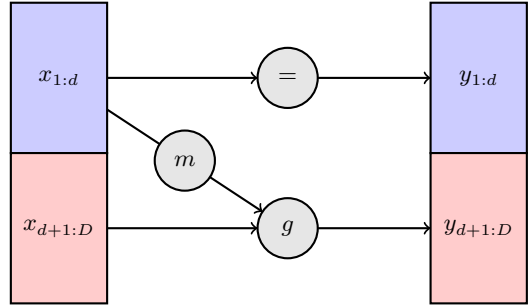

RQ-RSCs are bijections that are composed of coupling layers \citep{dinh2015, dinh2017}.
A coupling layer partitions the data, $x \in \R^D$, into two sets, $x_{1:d}$ and $x_{d+1:D}$.
The first set is then used to transform the second set:
\begin{align}
    \begin{split}
    y^{}_{1:d} &= x^{}_{1:d} \\
    y^{}_{d+1:D} &= g(x^{}_{d+1:D}; m(x^{}_{1:d})),
    \end{split}
\end{align}
where $g : \R^{D-d} \times \R^d \to \R^{D-d}$ is an invertible \emph{coupling law}, and $m$ is a \emph{coupling function} defined on $\R^d$.
This is illustrated in Figure~\ref{fig:coupling}.
The advantage of this structure is that the Jacobian is triangular,
\begin{align}
    \frac{\partial y}{\partial x} =
    \begin{pmatrix}
         I_d & 0 \\
         \frac{\partial y^{}_{d+1:D}}{\partial x^{}_{1:d}}
         & \frac{\partial y^{}_{d+1:D}}{\partial x^{}_{d+1:D}}
    \end{pmatrix},
\end{align}
where $I_d$ is the $d \times d$ identity matrix.
In particular, the Jacobian determinant is
\begin{align}
    \det \frac{\partial y}{\partial x} = \det \frac{\partial y^{}_{d+1:D}}{\partial x^{}_{d+1:D}}.
\end{align}
Furthermore, the inverse can be calculated as
\begin{align}
    \begin{split}
    x^{}_{1:d} &= y^{}_{1:d} \\
    x^{}_{d+1:D} &= g^{-1}(y^{}_{d+1:D}; m(x^{}_{1:d})),
    \end{split}
\end{align}
Notice that neither inverting a coupling layer $g$, nor calculating the Jacobian determinant, requires inverting or taking derivatives of the coupling function $m$, which can thus be arbitrarily complex.

The obvious limitation of a coupling layer is that only a subset of the data dimensions are transformed.
This is overcome by stacking multiple coupling layers in succession, and switching which variables belong to which partition.
In practice, this is achieved by interspersing coupling layers with bijections that shuffle the dimensions of $x$.
These shuffling bijections are trivially inverted and have a Jacobian determinant of one.

In a general coupling layer $g$, there are a variety of coupling laws $m$ one can use.
RQ-RSC's use Rational-Quadratic Neural Spline Coupling \citep{durkan2019}.
As the name  suggests, the coupling law $g$ is a set of rational-quadratic splines.
In particular, $g_i: [-B, B] \to [-B, B]$ for each dimension $i$ of $x_{d+1:D}$, where $g_i$ is a piecewise combination of $K$ segments, and each segment is a rational-quadratic function.
The positions and derivatives of the knots that parameterize the splines are calculated using the coupling function $m$, which is a dense neural network applied to $x_{1:d}$.

The result is a bijection that achieves state-of-the-art performance and efficiency for forward modeling and density estimation \citep{kobyzev2020}, and are flexible enough to model complex distributions with multiple discontinuities and hundreds of modes.
In addition, they are easily adaptable for flows with periodic topology (Section~\ref{sec:periodic}).
For more details, see \citet{durkan2019}.

In this work, we stack Rational-Quadratic Neural Spline Couplings, with Rolling Layers between each -- a configuration we name Rational-Quadratic Rolling Spline Couplings (RQ-RSCs).
Rolling Layers shift the dimensions of $x$ by one place:
\begin{align}
    \mathrm{Roll}: [x_1, \dots , x_{D-1}, x_D] \to [x_D, x_1, \dots , x_{D-1}].
\end{align}
By constructing a stack with $D$ coupling layers, RQ-RSCs individually transform each of the $D$ dimensions of $x$ as a function of the other $D-1$ dimensions.
This allows the network to learn complex relationships between every subset of the dimensions.
In the limit of high spline resolution (i.e. $K \to \infty$), RQ-RSCs can model any differentiable, monotonic function on $[-B, B]^D$ and can thus model arbitrarily complex distributions in this region.
In practice, we find very good performance for diverse data sets with $K \approx 16$.

Note you can specify a different value of $K$ for each of the $D$ spline layers in order to individually control the resolution of each dimension.
Lowering $K$ typically results in a smoother distribution, while increasing $K$ increases the complexity the normalizing flow can capture, while also increasing computational and memory cost.

\subsubsection{Data processing bijections}
\label{sec:data-processing}

While RQ-RSCs perform the heavy lifting of mapping the data distribution $\px$ onto the latent distribution $\pu$, it is also convenient to define other bijections that perform useful operations such as pre- and post-processing.
We name these \emph{data processing bijections}.

For example, RQ-RSCs (and the RQ-NSCs on which they are based) are defined on the domain [-B, B], and thus will not transform samples outside this range.
It is therefore useful to define a \emph{Shift Bounds} bijection, which shifts the original range of each dimension to match the domain of the splines.
Note this shift must be set at training time, with the assumption that future test data will lie within the same bounds\footnote{While this sounds quite restrictive, neural networks are typically pretty bad at extrapolating beyond the bounds of the training set anyway.}.
You can choose a range wider than that covered by the training set if you wish to allow the flow to sample outside the range of the training set

For an example of building an application-specific data processing bijection, see the \emph{Color Transform} bijection defined in Section~\ref{sec:fwd-model}, which maps galaxy magnitudes to galaxy colors.
See section~\ref{sec:discrete} for data processing bijections that enable modeling of discrete data.

Instead of using these data processing bijections, you can of course manually pre-process the data before evaluating densities and post-process samples drawn from the normalizing flow.
However, by building pre- and post-processing directly into the bijection, you remove these extra steps from the workflow.
This reduces the complexity of working with the normalizing flow and ensures that the flow always ``remembers'' how to correctly perform these pre- and post-processing steps.

\subsection{Choosing a latent distribution}
\label{sec:latent}

In principle, with a sufficiently expressive bijection, the choice of latent distribution does not matter as long as it is a distribution in which you can easily sample and calculate densities.
However, in practice, bijections are limited in expressiveness, i.e. they cannot necessarily transform any arbitrary data distribution into any arbitrary latent distribution.

For example, the splines of RQ-RSCs only transform samples in the range [-B, B].
Sampling from a latent distribution with support outside this range will therefore result in strange outliers and incorrect boundary conditions.
One can apply a transformation to the latent samples before they are fed into the RQ-RSC to ensure that they lie within the support of the splines, but it is simpler to use a compact latent distribution whose support matches that of the spline layers.
A simple choice would be the uniform distribution over [-B, B].

Additionally, as no bijection is perfect, the structure of the latent distribution will not be completely erased in the translation from latent to data distribution.
Thus, the latent distribution can be viewed as a prior or inductive bias on samples from the data distribution \citep{jaini2020}.
It is therefore advantageous to select a latent distribution whose features match some of the structure in the data.

A latent distribution that can achieve both desiderata is the Beta distribution, i.e. $u \sim \mathrm{Beta}(\alpha, \beta)$, where $\alpha, \beta > 0$ are learnable parameters\footnote{In practice, it is easier to learn $\log\alpha$ and $\log\beta$ to ensure that $\alpha, \beta > 0$.}.
This distribution is compact, and by varying $\alpha$ and $\beta$ this distribution can take on a wide variety of shapes with different means, skews, and kurtoses, allowing the inductive bias of the prior to adapt to structure in the data during training.
However, the Beta distribution is defined on the domain $[0, 1]$, while RQ-RSCs are defined on $[-B, B]$.
It is therefore more convenient to use a modified Beta distribution, which we name the \emph{Centered Beta distribution}:
\begin{align}
    \text{CentBeta}(u | \alpha, \beta, B) = 2B\left(\text{Beta}(u|\alpha, \beta) - \frac{1}{2}\right).
\end{align}

In general, as long as sampling and density evaluation are tractable, one can use any parameterization of the latent distribution that matches some desired structure in the data and learn the distribution parameters during training.
We give this generalization the name \emph{latent-adaptive flows} (LAFs; inspired by the Tail Adaptive Flows of \citealt{jaini2020}).
Learnable latent distributions can improve training loss, but require more care in training.

Note that while we discussed univariate distributions above, these considerations generalize easily to multiple dimensions.
Each of these distributions have multivariate generalizations that can be used when modeling higher-dimensional data.
The full multivariate latent distribution can also be assembled by taking the product of multiple univariate distributions\footnote{Note that while the latent variables will be independent, the data variables will still have correlations imprinted by the bijections.}.
This may even be desired if different dimensions of the data have different structure that you wish to encode in the latent distribution.

\subsection{Conditional flows}
\label{sec:conditional}

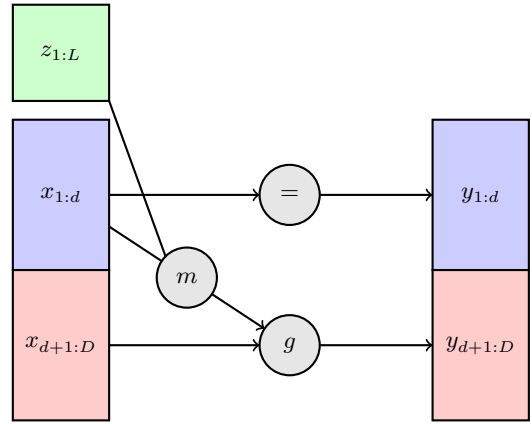
\begin{figure}
    \centering
    \begin{tikzpicture}[thick, outer sep=0]
        \node [draw, rectangle, minimum height=2cm, minimum width=1.28cm, fill=blue!20] (1) {$x^{}_{1:d}$};
        \node [draw, circle, minimum size=8mm, fill=gray!20, right = 2cm of 1] (2) {$=$};
        \node [draw, rectangle, minimum height=2cm, minimum width=1.28cm, fill=blue!20, right = 1.5cm of 2] (3) {$y^{}_{1:d}$};
        \draw[->] (1) -- (2);
        \draw[->] (2) -- (3);
        \node [draw, rectangle, minimum height=2cm, minimum width=1.28cm, fill=red!20, below = 0 of 1] (4) {$x^{}_{d+1:D}$};
        \node [draw, circle, minimum size=8mm, fill=gray!20, right = 2cm of 4] (5) {$g$};
        \node [draw, rectangle, minimum height=2cm, minimum width=1.28cm, fill=red!20, below = 0 of 3] (6) {$y^{}_{d+1:D}$};
        \draw[->] (4) -- (5);
        \draw[->] (5) -- (6);
        \draw[->] (1) -- (5) node [midway, draw, circle, fill=white, minimum size=8mm, fill=gray!20] (7) {$m$};
        \node [draw, rectangle, minimum height=1.28cm, minimum width=1.28cm, fill=green!20, above = 0.25 of 1] (8) {$z^{}_{1:L}$};
        \draw[-] (8.south east) -- (7.north west);
    \end{tikzpicture}
    \caption{
        Diagram of a \emph{conditional} coupling layer.
        The first partition, $x^{}_{1:d}$, is passed through the layer unchanged.
        The second partition, $x^{}_{d+1:D}$, is transformed by the coupling law $g$, which is parameterized by the coupling function $m$ applied to the first partition \emph{and} the conditional variables $z^{}_{1:L}$.
        The conditional variables are \emph{never} altered by the flow.
    }
    \label{fig:conditional-coupling}
\end{figure}

The bijections and latent distributions discussed above can be easily adapted to directly learn conditional probability distributions:
you only need to make the replacement $f(x) \to f(x;z)$, where $z$ is a vector of conditions \citep{winkler2019}.
This is illustrated in Figure~\ref{fig:conditional-coupling}, which is a modification of Figure~\ref{fig:coupling} to include the input of conditional variables to the coupling function $m$.
In practice, since $m$ is usually a neural network, this amounts to just appending the conditions $z$ to the inputs of the neural network.

While $p(x|z)$ is technically encoded within $p(x,z)$, which can be learned with a regular normalizing flow, directly modeling $p(x|z)$ with a conditional flow has a few benefits.
Training is typically faster, since the latent distribution has a smaller number of dimensions.
You can also draw samples of $x$ at fixed values of the conditions $z$, and you can calculate $p(x|z)$ without having to numerically calculate and divide by $p(z)$, which can be computationally expensive.

\subsection{Flows with periodic topology}
\label{sec:periodic}

The flows we have considered so far model data that live in $\R^n$.
This assumption is insufficient for modeling variables from spaces with non-Euclidean topology, e.g. positions on the sky.
While progress has been made on building flows for general topologies (e.g. \citealt{gemici2016} and \citealt{falorsi2019}), we will focus on building flows on the sphere, $S^2$, as this is the case most relevant in astronomy.
We will see that by carefully choosing the latent space, we can construct flows with periodic topology with minimal additional effort \citep{rezende2020}.

Positions on the sphere are specified by two angles\footnote{
We use the convention where $\theta$ and $\phi$ are the zenith and azimuthal angles, respectively.
},
$\theta$ and $\phi$, the latter of which is periodic.
By mapping $\theta$ to $\cos\theta$, we map the sphere to a cylinder\footnote{
This map can be explicitly constructed via an embedding in $\R^3$.
Technically, the map is not defined for $\theta \in \{0, \pi\}$, however as this set has zero measure, it can be safely ignored.}:
$S^2 \to [-1,1] \times S^1$ (i.e. the Cartesian product of an interval and a circle).
In other words, we can transform $\cos\theta$ with a Euclidean flow, as long as we ensure that the flow bounds samples to the range $[-1, 1]$.
However, the $S^1$ piece, $\phi$, has a periodic topology and must be handled more carefully.

First, we will address transformations of $\cos\theta$.
The only constraint we must impose is that samples of $\cos\theta$ must lie in the range $[-1, 1]$.
Fortunately, RQ-RSCs are bounded, mapping a range in $u$ to the same range in $x$.
Thus, if we pick a latent distribution with compact support in $[-1, 1]$, samples of $\cos\theta$ are guaranteed to lie in the same range, as long as we set the range of the RQ-RSC $B$ = 1.

Next we will address transformations of $\phi$.
For $f$ to be a differentiable bijection on the circle, $S^1$, it is sufficient that $f$ obey the following constraints:
\begin{align}
    f(0) &= 0 \\
    f(2\pi) &= 2\pi \\
    \nabla f(0) &= \nabla f(2\pi) \label{eq:df=df} \\
    \nabla f(\phi) &> 0.
\end{align}
The first two constraints ensure continuity of $f$ by designating $\phi=0$ as a fixed point, and the third constraint ensures continuity of $\nabla f$ at that fixed point.
While the designation of $\phi=0$ as a fixed point is an unnecessary restriction on $f$, any differentiable bijection on the circle has at least one fixed point up to a phase change, and so this restriction does not actually restrict the expressiveness of $f$.
The fourth restriction ensures monotonicity, which guarantees invertibility.

If we make the phase change $\phi \to \phi - \pi$ so that our angles $\phi \in [-\pi, \pi]$, a RQ-NSC with $B=\pi$ automatically fulfills all four constraints.
In fact, regular RQ-NSC's impose the further condition
\begin{align}
    \nabla f(-\pi) = \nabla f(\pi) = 1 \label{eq:df=1}
\end{align}
to match an identity transform for inputs outside of the range $[-\pi, \pi]$.
By choosing a latent distribution with compact support in the range $[-\pi, \pi]$, we ensure that no samples will lie outside the range of the splines, and so we can relax the boundary condition of Equation~\ref{eq:df=1} in favor of the boundary condition in Equation~\ref{eq:df=df}.
Spline transforms with this relaxed boundary condition are named \emph{Circular Splines} by \citet{rezende2020}.

The circular spline construction above is easily generalized to n-spheres and n-tori: $S^n \to [-1, 1]^{n-1} \times S^1$ and $T^n \to (S^1)^n$ (see \citealt{rezende2020} for more details).
We can model the joint distribution of periodic and non-periodic variables with RQ-RSCs simply by choosing appropriate bounds $B$ for each dimension, and by swapping boundary condition~\ref{eq:df=1} for condition~\ref{eq:df=df} for any periodic dimensions.

\subsection{Modeling discrete variables}
\label{sec:discrete}

In addition to the continuous variables described above, normalizing flows can also be used to model discrete variables.
This can be achieved by ``dequantizing'' the discrete dimensions of the data, which can then be mapped onto continuous latent distributions using regular continuous bijections.
Dequantization consists of adding some kind of continuous noise to the discrete dimensions, transforming them into continuous dimensions.
When sampling from the flow, you simply do the opposite, and ``quantize'' the discrete dimensions after applying all of the regular bijections, mapping the noisy, continuous variables onto their discrete counterparts.

A common method for dequantization is uniform dequantization, in which random uniform noise in the range (0, 1) is added to the discrete dimensions.
The corresponding quantization applied while sampling from the flow consists of applying the floor function to the dequantized dimensions, mapping these samples onto the nearest integer less than the sampled value.
More sophisticated dequantization schemes use variational inference or even another normalizing flow to determine the noise distributions, which improves results by smoothing the discontinuities between neighboring discrete values.
See \citet{ho2019} \citet{hoogeboom2020} for more details.

While the dequantizers are not technically bijections, they can be treated as data processing bijections and be chained together with the other bijections in your normalizing flow.

\section{PZFlow}
\label{sec:pzflow}

PZFlow is a Python package for building normalizing flows, with a focus on features useful for forward modeling and density estimation for tabular data.
Data is handled in Pandas DataFrames \citep{pandas}, while the normalizing flows are implemented in Jax \citep{jax}, which allows for efficient, parallelizable, GPU-enabled calculations for very large data sets.
The code is easily installable from the Python Package Index\footnote{\url{https://pypi.org/project/pzflow/}} (PyPI) and is hosted on Github\footnote{\url{https://github.com/jfcrenshaw/pzflow}}.
The documentation\footnote{\url{https://jfcrenshaw.github.io/pzflow/}} includes tutorial notebooks demonstrating the features mentioned in this paper on different example problems.

The rest of this paper will focus on forward modeling a photometric galaxy catalog, and photo-z inference.
Section~\ref{sec:galaxy-catalog} uses PZFlow to forward model a galaxy catalog, including photometry, spectroscopic redshifts (spec-z's), true photo-z posteriors, ellipticities, and sizes.
Section~\ref{sec:photo-z} uses PZFlow for photo-z estimation, demonstrating the power of PZFlow as a density estimator, including numerous useful features for photo-z estimation.

In addition to the examples in this paper, PZFlow has already being used in various other DESC projects:
\begin{itemize}
    \item \citet{malz2021} used PZFlow to build a metric for observing strategy optimization based on information theory;
    \item \citet{stylianou2022} used PZFlow to forward model galaxy data with true redshift posteriors in order to evaluate the impact of survey incompleteness and spec-z errors on photo-z estimation;
    \item \citet{lokken2022} used PZFlow to smooth high-redshift artifacts in simulations of host galaxies for supernovae and other transients;
    \item \citet{moskowitz2024} used PZFlow to explore data augmentation for photo-z spectroscopic training sets.
\end{itemize}

\section{Forward Modeling a Galaxy Catalog}
\label{sec:galaxy-catalog}

In this section, we use PZFlow to forward model a photometric galaxy catalog for the Vera Rubin Observatory's Legacy Survey of Space and Time (LSST; \citealt{ivezic2019}).
The advantage of using a catalog generated from a normalizing flow is that we have direct access to the probability distribution from which the data is drawn, enabling us to calculate true values for derived statistical products, such as the true photo-z redshift posterior for each galaxy.

In Section~\ref{sec:fwd-model} we construct a normalizing flow to model the galaxy redshifts and photometry and generate a new simulated catalog.
In Section~\ref{sec:true-posteriors}, we calculate true redshift posteriors for the new catalog.
In Section~\ref{sec:fwd-model-conditional} we build a conditional flow to add additional galaxy properties to the catalog.

\subsection{Forward modeling redshifts and photometry}
\label{sec:fwd-model}

\begin{figure*}[t]
    \script{forward_model/plot_main_galaxy_corner.py}
    \begin{centering}
        \includegraphics{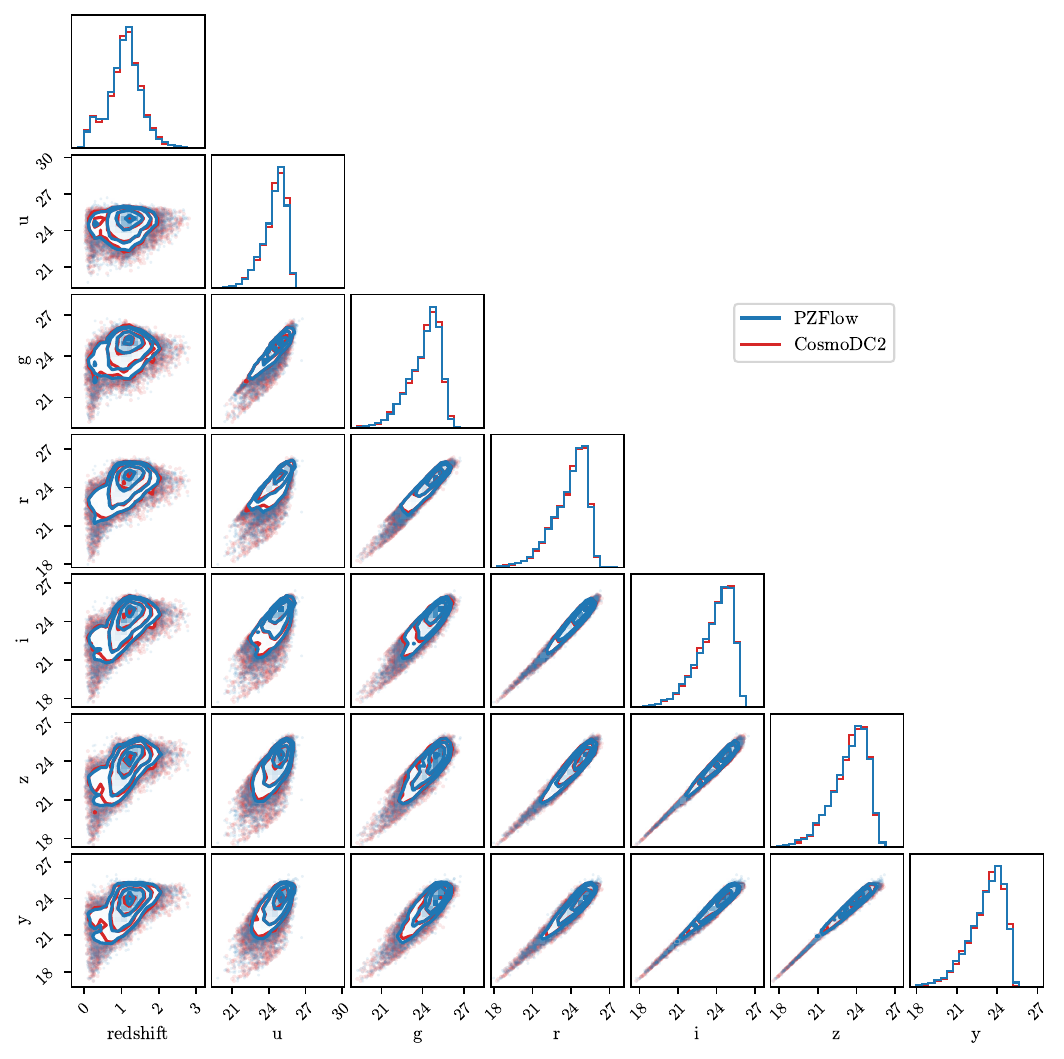}
        \caption{
            Distribution of true redshifts and noisy photometry from the CosmoDC2 test set, compared to a sample drawn from the distribution learned by PZFlow.
            The close overlap of every pair-wise distribution demonstrates that PZFlow has learned the distribution in CosmoDC2 with high fidelity.
        }
        \label{fig:main-corner}
    \end{centering}
\end{figure*}

\begin{figure*}[t]
    \script{forward_model/plot_smooth_color_distribution.py}
    \begin{centering}
        \includegraphics{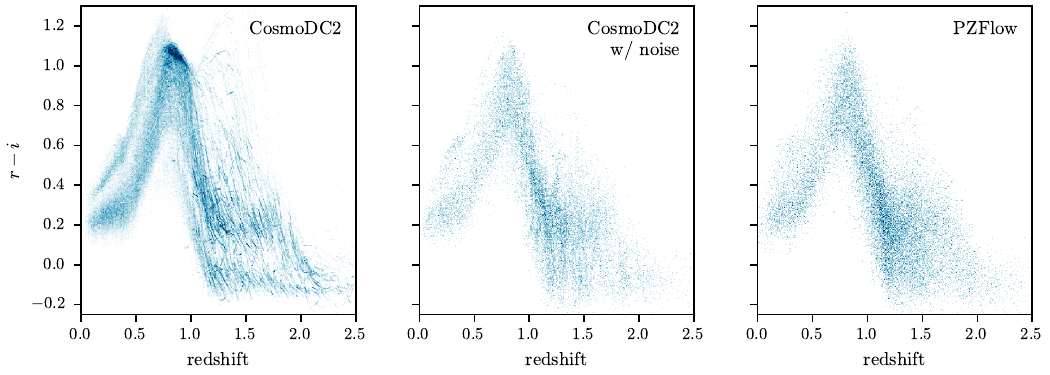}
        \caption{
            Comparison of the $r-i$ vs redshift distribution for galaxy samples from CosmoDC2 without photometric noise (left), CosmoDC2 with photometric noise (middle), and from the normalizing flow (right).
            High-redshift galaxies in CosmoDC2 lie along discrete tracks in redshift-color space.
            Adding photometric noise somewhat smooths but does not totally remove these tracks.
            PZFlow produces a catalog with a smooth redshift-color distribution.
        }
        \label{fig:smooth-color-dist}
    \end{centering}
\end{figure*}

To create a generative model of galaxy redshifts and photometry, we use the true redshifts and $ugrizy$ magnitudes from the CosmoDC2 simulation \citep{dc2, cosmodc2} of the LSST Dark Energy Science Collaboration (DESC).
We add photometric errors to the true $ugrizy$ magnitudes using the 10-year-depth LSST extended-source error model of our PhotErr package (see Appendix~\ref{app:error-model}), and selected galaxies with a signal-to-noise ratio (SNR) greater than 5 in the $i$ band.
Of these, we randomly selected $10^6$ galaxies and split them into training and test sets consisting of 80\% and 20\% of the galaxies, respectively.
We then train a normalizing flow to learn the distribution $p(z, \mathbf{\hat{m}})$, where $z$ is the true redshift, and $\mathbf{\hat{m}}$ is the vector of noisy magnitudes in the LSST bands.

For the latent distribution we use a 7 dimensional Uniform distribution over the range $[-5, 5]$\footnote{The choice of 5 was arbitrary. Any other positive value would work just as well.}.
To map the data onto the latent distribution, we use the following bijection:
\begin{align}
    f = \text{RQ-RSC} \circ \text{Shift Bounds} \circ \text{Color Transform}.
\end{align}
We will explain each layer of the bijection in the order they are applied to the input data.

The first layer of the bijection is the Color Transform, a data processing bijection designed specifically for this task.
The Color Transform converts galaxy magnitudes to colors, but keeps the $i$ band magnitude as a proxy for the apparent luminosity:
\begin{multline}
    \text{Color Transform} : (\text{redshift},\, u,\, g,\, r,\, i,\, z,\, y) \to \\
    (\text{redshift},\, i,\, u-g,\, g-r,\, r-i,\, i-z,\, z-y).
\end{multline}
This layer is useful as galaxy redshifts correlate more directly with galaxy colors than galaxy magnitudes.

The next layer, Shift Bounds, is the data processing bijection defined in Section~\ref{sec:data-processing}, which maps the range of the data onto the support of the RQ-RSC.
Note that since Shift Bounds is on the ``other side'' of the Color Transform, we need to map the ranges of the colors $u-g$, $g-r$, etc. onto the support of the splines, instead of the original magnitudes.

The final layer is an RQ-RSC, described in detail in Section~\ref{sec:rq-rsc}.
This layer performs the heavy lifting of transforming the data distribution into the uniform latent distribution.
We use $D=7$ layers to transform all 7 dimensions of our data, and set $B=5$ to match the support of the latent distribution.
We use the coupling function (a feedforward neural network with two hidden layers of 128 neurons) described in \citet{durkan2019}.
We use $K=16$ spline knots.
This number was chosen to be large enough to capture the complexity of the data, but small enough so that the flow smooths over the discrete tracks in Figure~\ref{fig:smooth-color-dist} (more on this below).

After training the flow (see Appendix~\ref{app:training-details}), we assess the results by drawing $10^4$ galaxies from the trained flow, and plotting their distribution against  $10^4$ galaxies from the test set (Figure~\ref{fig:main-corner}).
We see the normalizing flow has done an excellent job of reproducing the distribution of galaxies in CosmoDC2, without any unusual artifacts or outliers.
In addition, Figure~\ref{fig:smooth-color-dist} compares the distribution of galaxy $r-i$ vs redshift.
High-redshift galaxies in the CosmoDC2 simulation lie on discrete tracks in this space due to the way galaxies were assigned to a discrete set of SED templates during simulation.
These tracks are easily visible in the left panel.
Adding photometric errors somewhat smooths the distribution, but close inspection reveals there is still granularity in the distribution of high-redshift galaxies.
The right panel shows the distribution produced by PZFlow, which has a smooth color distribution, even at high redshift.
We note that these results were obtained without any extensive hyperparameter search, and that very similar (slightly worse results) are obtained without the \texttt{ColorTransform} bijection, demonstrating the flexibility of the method to adapt to unseen data sets.

With this normalizing flow, we have an efficient, probabilistic CosmoDC2 emulator that models a smooth color-redshift distribution up to redshift 3.
We generate a catalog by sampling $10^4$ galaxies from the flow, each with noisy photometry and a true redshift.
Importantly, since we have access to the probability distribution from which the galaxies were generated, we can calculate true redshift posteriors for each galaxy.
This is the subject of the next section.

\subsection{Calculating true posteriors}
\label{sec:true-posteriors}

\begin{figure}[t]
    \script{forward_model/plot_posteriors.py}
    \begin{centering}
        \includegraphics{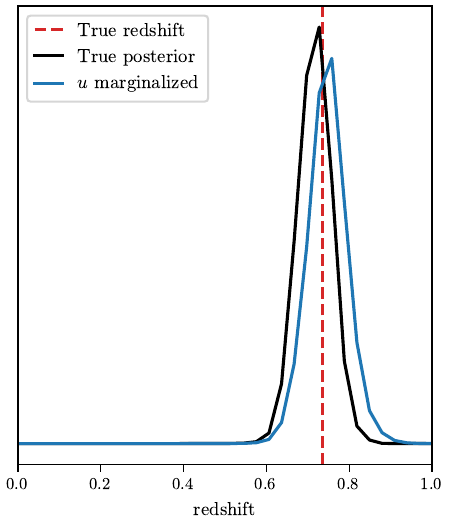}
        \caption{
            Example of redshift posteriors for a galaxy simulated with PZFlow.
            The true redshift of the galaxy is marked by the vertical dashed red line, and the true redshift posterior for the galaxy is drawn in black.
            Calculating the posterior while marginalizing over the $u$ band magnitude yields the posterior in blue.
            Note that the $u$ band marginalization is only approximate, but increasing the resolution of the grid of $u$ band values causes the resultant posterior to converge.
        }
        \label{fig:posteriors}
    \end{centering}
\end{figure}

\begin{figure*}[t]
    \script{forward_model/plot_conditional_galaxy_corner.py}
    \begin{centering}
        \includegraphics{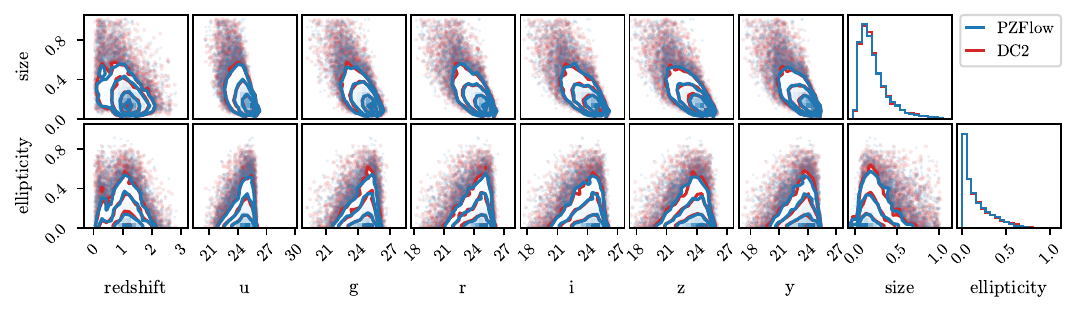}
        \caption{
            Conditional distributions of the ellipticity and size of the galaxies in the CosmoDC2 test set compared to the distribution learned by PZFlow.
            The close overlap of every pair-wise distribution demonstrates that PZFlow has learned the distribution in CosmoDC2 with high fidelity.
        }
        \label{fig:conditional-corner}
    \end{centering}
\end{figure*}

Since we have direct access to the probability distribution from which the photometry and redshifts are drawn, using Equation~\ref{eq:px}, we can analytically calculate the true redshift posterior for each galaxy: $p(z|\mathbf{\hat{m}})$ where $\mathbf{\hat{m}}$ is the vector of noisy galaxy magnitudes.
We note this is not an estimate, like would be returned by a photo-z estimator, but rather the simulated truth, obtained from the model that generated the photometry and redshifts in the first place.
Of course the resolution of the posterior is limited to the choice of redshift grid.

When calculating these posteriors, for each galaxy we can also marginalize over the magnitudes in any missing bands.
Imagine, for a galaxy, that we partition the vector of magnitudes $\mathbf{\hat{m}}$ into an observed set $\mathbf{\hat{m}}_0$ and a missing set $\mathbf{\hat{m}}_\text{x}$.
We can marginalize over the missing magnitudes when calculating the posterior
\begin{align}
    p(z|\mathbf{\hat{m}}_0) = \frac{1}{p(\mathbf{\hat{m}}_0)} \int p(z, \mathbf{\hat{m}}_0, \mathbf{\hat{m}}_\text{x}) d\mathbf{\hat{m}}_\text{x},
\end{align}
which can be calculated by evaluating $p(z, \mathbf{\hat{m}})$ on a grid of $z$ and possible values of $\mathbf{\hat{m}}_\text{x}$, summing over $\mathbf{\hat{m}}_\text{x}$ to yield $p(z, \mathbf{\hat{m}}_0)$, and normalizing with respect to redshift to yield $p(z|\mathbf{\hat{m}}_0)$.
PZFlow possess a flexible method for performing this marginalization: the grid for each band in $\mathbf{\hat{m}}_\text{x}$ can be a function of other galaxy properties (e.g. the observed magnitudes, $\mathbf{\hat{m}}_0$).

You may wish to marginalize over all values of $\mathbf{\hat{m}}_\text{x}$ if the galaxy was not observed in those bands.
This may occur, for example, when simulating a joint Euclid-LSST catalog \citep{euclid}, as not all galaxies will have photometry from both.
You may also wish to marginalize over all values beyond the limiting magnitudes to simulate a galaxy that was observed but not detected in the corresponding bands.
This might occur, for example, in the low wavelength bands of Lyman-dropout galaxies observed by LSST.

Note this marginalization is only approximate, and therefore weakens our ability to refer to these as ``true'' redshift posteriors.
However, increasing the resolution of the grid used for $u$ band marginalization causes the resultant posteriors to converge.
Thus, we believe it is still appropriate to treat these marginalized posteriors as the truth for the purpose of photo-z validation.

Redshift posteriors for an example galaxy are displayed in Figure~\ref{fig:posteriors}.
The black posterior is calculated using the full set of galaxy magnitudes.
The true redshift, marked by the vertical red line, nearly coincides with the mode of this posterior.
The blue posterior has been calculated while marginalizing over the $u$ band.
Throwing away the information in the $u$ band slightly broadens the posterior and shifts it toward higher redshifts.

Calculating these posteriors enables direct comparison of true redshift posteriors with the redshift posteriors estimated by photo-z estimators.
This is important, as modern cosmology analyses are beginning to increasingly rely on full redshift posteriors \citep{mandelbaum2008,newman2022}.
\citePZt showed that popular metrics for evaluating photo-z estimators using ensembles of photo-z posteriors can be misleading, and are not well suited to the needs of precision cosmology.
PZFlow catalogs with true redshift posteriors provide a path forward by enabling the evaluation of photo-z estimators on a per-posterior basis.

\subsection{Additional properties with conditional flows}
\label{sec:fwd-model-conditional}

\begin{figure*}[t]
    \script{photo-z/plot_ensemble_posteriors.py}
    \begin{centering}
        \includegraphics{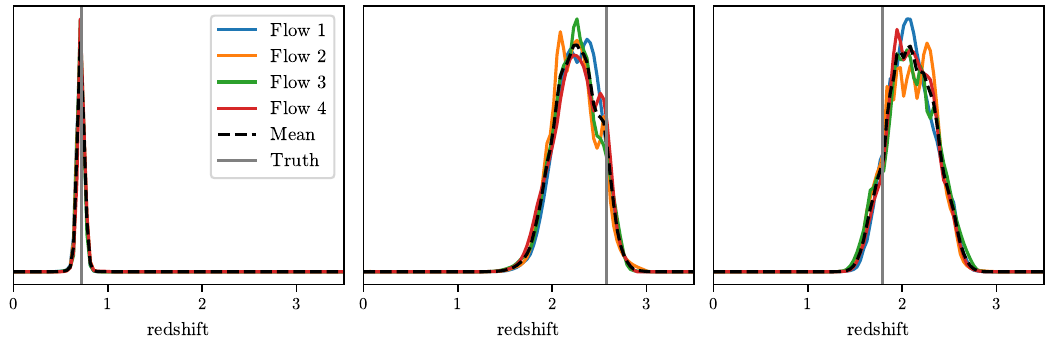}
        \caption{
            The ensemble of posteriors for three example galaxies.
            Flows 1-4 label the individual posteriors produced by each of the flows that make up the ensemble.
            The dashed black line is the mean of these individual posteriors and is the value used by the ensemble.
            The vertical gray line labeled ``Truth'' denotes the true redshift of the galaxy.
            Averaging the posteriors from each flow in the ensemble approximately marginalizes over the neural network parameters, and smooths over the small-scale variations found in the posterior from any individual flow.
            The first panel is a galaxy with a narrow and unimodal redshift posterior, while the next two panels demonstrate broad, multimodal posteriors, which is typical of galaxies in the range $1.5 < z < 2.6$.
        }
        \label{fig:ensemble-posteriors}
    \end{centering}
\end{figure*}

In addition to the galaxy magnitudes and redshifts modeled above, we wish to include other galaxy properties in the catalog, such as galaxy size and ellipticity.
In principle, we could have included these variables in the original normalizing flow.
However, we did not want the true redshift posteriors to be conditioned on these variables, as most photo-z estimators only use galaxy photometry.
Therefore, we will build a second flow that models these additional values conditioned on the galaxy redshift and magnitudes.
In other words, we are modeling the full joint distribution via the decomposition
\begin{align}
    p(z, \mathbf{\hat{m}}, s, e) = p(z, \mathbf{\hat{m}}) \cdot p(s, e | z, \mathbf{\hat{m}}),
    \label{eq:chain-rule}
\end{align}
where $s$ is the size (the half-light radius in arcseconds) and $e$ is the ellipticity.
The first distribution on the right hand side of Equation~\ref{eq:chain-rule} is modeled by our original flow, and the second distribution will be modeled using the new conditional flow.
While we have only chosen to model these additional two properties, any other values you desire can be similarly modeled.

For the latent distribution, we again use a Uniform distribution over the range $[-5, 5]$.
For the bijection, we use
\begin{align}
    \begin{split}
        f =& ~ \text{RQ-RSC} \circ \text{Shift Bounds}. \\
    \end{split}
\end{align}
The RQ-RSC acts on the two dimensional space of size and ellipticity, but also takes the galaxy redshift and magnitudes as inputs (see the conditional variables in green in Figure~\ref{fig:conditional-coupling}).
The redshifts and magnitudes are transformed to have zero mean and unit variance before being input to the neural network\footnote{These variables are standard scaled instead of mapped onto the domain [-5, 5], because the neural network that parameterizes the splines has no limit on inputs, unlike the splines themselves, which are limited to the range [-5, 5].} that parameterizes the splines.
Aside from the change in inputs, the RQ-RSC has the same settings as listed for the previous normalizing flow.

After training the flow (see Appendix~\ref{app:training-details}), we sample a size and ellipticity for each galaxy in the PZFlow catalog created in the previous section (conditioned on the redshift and noisy magnitudes), and plot the distribution of these features against the distribution in the test set (Figure~\ref{fig:conditional-corner}).
Once again, we see the normalizing flow does a good job of emulating the CosmoDC2 galaxy distribution.
We note that, if desired, the TARP test of \citet{lemos2023} can provide a quantitative test of the fidelity of the conditional flow. 

The final simulated catalog consists of $10^4$ galaxies, each with a redshift, noisy $ugrizy$ magnitudes, a true photo-z posterior, a size, and an ellipticity.
We use the magnitudes, size, and ellipticity to estimate the photometric errors using the 10-year-depth LSST extended-source error model of PhotErr.
This small catalog was generated for visualization purposes, but the normalizing flows can be used to generate catalogs of arbitrarily large size.
In particular, in only a few minutes, one can generate new catalogs or augment existing catalogs with millions of galaxies.
This is substantially faster than re-running large scale simulations like CosmoDC2.

\section{Photometric Redshift Estimation}
\label{sec:photo-z}

In addition to forward modeling, normalizing flows are powerful and flexible models for density estimation.
This makes them useful tools for estimating posterior distributions for galaxy properties, conditioned on observed features of the galaxy.
In this section, we demonstrate this by applying PZFlow to photo-z estimation for the simulated catalog from the previous section.

\subsection{Training an Ensemble for photo-z estimation}

When forward modeling in Section~\ref{sec:galaxy-catalog}, we wanted a realistic model that captured the relevant correlations between galaxy photometry, redshift, shape, and size.
However, when estimating redshifts, we do not simply want a realistic model, but rather a model that matches our specific galaxy sample as closely as possible.

When training deep learning models, the huge parameter space contains many different solutions, corresponding to different local minima in the parameter space.
In the forward modeling application, we were content with finding a good local minimum, but in this application, we want to marginalize over the different potential models.

A full marginalization over the model parameters would be too computationally expensive, so instead we approximate this marginalization using an ensemble of normalizing flows.
In other words, we train multiple normalizing flows under identical conditions, using different random initializations of the model parameters.
This allows the optimization algorithm to explore different basins of attraction in the parameter space.
In the machine learning literature, this is known as a Deep Ensemble \citep{lakshminarayanan2017}, and is a popular method for approximate bayesian marginalization \citep{wilson2020,fort2020}.

We train an ensemble of 4 normalizing flows, each with the same architecture and training schedule as the regular flow described in Section~\ref{sec:galaxy-catalog}.
With PZFlow, this is as simple as swapping \texttt{FlowEnsemble} for \texttt{Flow} in the code.

For the training set, we use 100,000 galaxies from the catalog created in Section~\ref{sec:galaxy-catalog}.
Each galaxy in the training set has a true redshift and observed noisy magnitudes in the $ugrizy$ bands, with corresponding photometric errors.
To account for the photometric error, at the start of each training epoch, we resample the training set from the photometric error distributions.
In other words, each epoch, for each galaxy, we sample
\begin{align}
    \mathbf{m} \sim p(\mathbf{\hat{m}}, \mathbf{\sigma_m}),
\end{align}
where $\mathbf{\hat{m}}$ are the observed magnitudes with photometric errors $\mathbf{\sigma_m}$, and $p(\mathbf{\hat{m}}, \mathbf{\sigma_m})$ is a Gaussian in flux space.
This allows our ensemble of flows to approximate the distribution $p(z, \mathbf{m})$, where $\mathbf{m}$ is the vector of true magnitudes for the galaxy.
For more details on training the ensemble, see Appendix~\ref{app:training-details}.

\subsection{Estimating posteriors}

After training, we use each flow in the ensemble to estimate the redshift posterior by marginalizing over the photometric errors:
\begin{align}
    p(z| \mathbf{\hat{m}}, \mathbf{\sigma_m}) \propto \int p(z, \mathbf{m}) \, p(\mathbf{m}| \mathbf{\hat{m}}, \mathbf{\sigma_m}) \, d\mathbf{m},
\end{align}
which is estimated by sampling $\mathbf{m} \sim p(\mathbf{\hat{m}}, \mathbf{\sigma_m})$ and averaging $p(z, \mathbf{m})$ over these samples.
We then average the $p(z, \mathbf{m})$ from each flow, and normalize with respect to the redshift grid.
This provides a redshift posterior for each galaxy.

Posteriors for three galaxies can be seen in Figure~\ref{fig:ensemble-posteriors}.
Each flow produces a PDF which may contain slightly different features in each case.
By averaging over the individual posteriors, we select for features that are common between models, while smoothing over features that are present in only a single model.
The first example galaxy in Figure~\ref{fig:ensemble-posteriors} is at $z < 1$, and all flows in the ensemble return essentially the same narrow redshift posterior.
This is typical for low-redshift galaxies whose photo-z's are relatively well constrained by LSST photometry.

The other two example galaxies, however, are in the $1.5 < z < 2.6$ redshift range, where the Balmer Break (at $\sim 4000$~\AA) has redshifted out of LSST's wavelength coverage, while the Lyman Limit (at $912$~\AA) has not yet redshifted into LSST's wavelength coverage.
As a result, these posteriors are much broader and less well constrained.
For these two galaxies, the flows in the ensemble return posteriors with different small-scale variations, and the best-estimate redshift (i.e. the mode of the redshift posterior) varies by as much as 0.5 for each set of posteriors.
Marginalizing over the individual posteriors smooths over these variations.
We can also treat the ensemble of posteriors as a distribution over possible posteriors, which will allow for more consistent error calibration in cosmological analyses \citep{zhang2023}.

\subsection{Photo-z metrics}

\begin{figure}[t]
    \script{photo-z/plot_pz_point_estimates.py}
    \begin{centering}
        \includegraphics{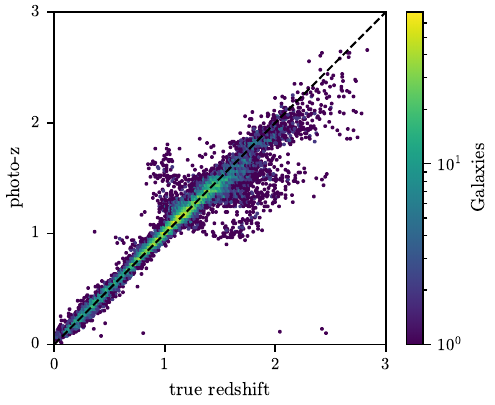}
        \caption{
            Photo-z point estimates (maximum a posteriori) vs true redshift for galaxies in the test set.
        }
        \label{fig:point-estimates}
    \end{centering}
\end{figure}

\begin{figure*}[t]
    \script{photo-z/plot_binned_metrics.py}
    \begin{centering}
        \includegraphics{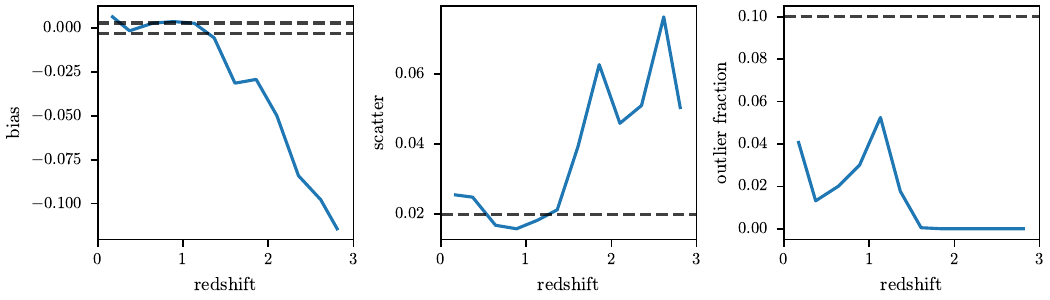}
        \caption{
            The bias, scatter, and outlier fraction of the photo-z point estimates as a function of true galaxy redshift.
            The dashed black lines represent the limits for LSST cosmology as stated in the LSST DESC SRD \citep{descSRD}.
            The bias must be between these dashed lines, while the scatter and outlier fraction must be below the dashed lines.
            You can see that PZFlow meets the bias and scatter requirements up to redshift $\sim$ 1.5 (the scatter is high at low redshifts due to the relative small number of low-redshift galaxies in our training set).
            PZFlow meets the outlier requirements for all redshifts.
            We note that individual redshifts do not actually need to meet the bias requirement as long as the bias can be well calibrated via some other source, e.g. galaxy clustering.
        }
        \label{fig:binned-metrics}
    \end{centering}
\end{figure*}

\begin{figure}[t]
    \script{photo-z/plot_pit_histogram.py}
    \begin{centering}
        \includegraphics{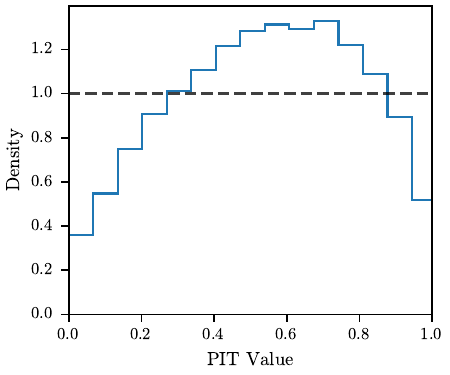}
        \caption{
            The probability integral transform (PIT) histogram for PZFlow photo-z posteriors.
            The PIT characterizes the calibration of the estimated posteriors, with the horizontal black line indicating perfect calibration.
        }
        \label{fig:pit-histogram}
    \end{centering}
\end{figure}

In this section, we evaluate the performance of PZFlow using common photo-z metrics.
Note these metrics are optimistic in the sense that the training set is representative of the test set, which is usually not the case in modern cosmology applications.

The most common metrics for photo-z estimation concern photo-z point estimates, which are a compression of the photo-z posterior to a single redshift estimate (e.g., \citealt{hildebrandt2010,sanchez2014}).
We make the common choice of selecting the mode of the posteriors\footnote{The mean redshift is a poor choice, since photo-z posteriors are often multimodal, and so the mean value can lie between two modes at a redshift with very small probability density.}.
We compute metrics of the quantity $\Delta z = (z_\text{phot} - z_\text{true}) / (1 + z_\text{true})$, where the denominator accounts for naturally greater uncertainties at high redshift.

Figure~\ref{fig:point-estimates} compares the photo-z point estimates to the true redshifts.
The point estimates for most galaxies lie along the diagonal, indicating strong performance.
There are the common photo-z ``wings'', indicating redshifts where important spectral features are transitioning between neighboring photometric bands.
This point estimate plot is comparable to other high-performance machine learning photo-z estimators when provided with representative training sets \citep{sanchez2014}.

Figure~\ref{fig:binned-metrics} shows the photo-z point estimate metrics from the LSST DESC Science Requirements Document (SRD; \citealt{descSRD}) as a function of true redshift.
The \emph{bias} is defined as the median of $\Delta z$; the \emph{scatter} is defined as $\text{IQR} / 1.349$, where IQR is the interquartile range of $\Delta z$; the \emph{outlier fraction} is defined as the fraction of galaxies for which $\Delta z$ is greater than three times the scatter.
The requirements from the SRD are plotted in black to provide a sense of scale.

Like many photo-z estimators, PZFlow performs well to a redshift of approximately 1.5 (the scatter is high at low redshifts due to the relatively small number of low-redshift galaxies in our training set).
At higher redshifts, our estimator does not meet the bias and scatter requirements, because there is very little training data in this redshift range.
We note however that for many cosmology applications, it is okay for the bias to exceed the required limits, as long as the bias can be well determined via some calibration process \citep{newman2015}.

Another common metric is the probability integral transform (PIT) (see e.g. \citePZa, \citealt{dey2022}), which is a histogram of the cumulative density function (CDF) of each posterior.
I.e., given an estimated posterior  $p(z| \mathbf{\hat{m}}, \mathbf{\sigma_m})$ and the true redshift $z_\text{true}$, the CDF is
\begin{align}
    \mathrm{CDF} = \int_0^{z_\text{true}} p(z| \mathbf{\hat{m}}, \mathbf{\sigma_m}) \, dz.
\end{align}
For perfectly calibrated posteriors, the CDF distribution (the PIT histogram) is uniform between $0$ and $1$.
This is because, for example, if the photo-z posteriors produced by an estimator are well calibrated you would expect the true redshifts of the galaxies to fall within the 50\% confidence intervals 50\% of the time.

The PIT histogram for our estimator is shown in Figure~\ref{fig:pit-histogram}.
Ideally, this histogram would be uniform and match the dashed horizontal line, which represents perfect calibration.
The fact that the histogram bulges at the center indicates our estimator is too conservative -- i.e. the posteriors it produces are too broad.
This can be explained by the fact that normalizing flows exhibit mode covering behavior (the opposite of the mode collapse seen in GANs; \citealt{salimans2016}).
In other words, because normalizing flows are trained by maximizing the likelihood of the training samples, they receive very high penalties for missing any modes in the data.
As a result, they tend to conservatively spread out their density, in order to avoid missing any modes.
This results in overly conservative posterior predictions.

The low values at the edges of the PIT histogram indicate the relative rarity of catastrophic outliers, which is also reflected in the far right panel of Figure~\ref{fig:binned-metrics}, where you can see our estimator meets the requirement on the outlier fraction at all redshifts.
There is also a slight rightward tilt.
This indicates a small negative bias, which reflects the intrinsic prior towards smaller redshifts, as this is where the majority of galaxies in the training set lie.
This negative bias is visible for high-redshift galaxies in the far left panel of Figure~\ref{fig:binned-metrics}.
Calibrating these posteriors, either via altering the training loss or post-processing the posteriors, is beyond the scope of this paper.
However this calibration could be achieved, for example, using the methods of \citet{dey2022}.

The previous metrics analyze photo-z performance for point estimates, which are insufficient for modern cosmology \citep{newman2022}, and for ensembles of posteriors, which is often misleading and not a good indicator of performance for science applications \citePZp.
The methods introduced in this paper enable the creation of galaxy catalogs for which each galaxy has a true redshift posterior, which will enable more comprehensive evaluation of photo-z estimators.
Full evaluation of photo-z estimators on a posterior-by-posterior basis is a major goal of the LSST DESC, and will be the focus of forthcoming work.

\section{Conclusion}
\label{sec:conclusion}

In this paper we introduced PZFlow, a Python package for probabilistic forward modeling of galaxy catalogs, and demonstrated how it will be used to assist the photo-z calibration efforts of the LSST DESC.
In particular, galaxies generated from a PZFlow model have a natural notion of a true photo-z posterior, to which the redshift posteriors estimated by photo-z algorithms can be directly compared.
This enables a more comprehensive evaluation of the posteriors produced by photo-z estimators that we expect will avoid the traps of ensemble-only metrics that were identified by \citePZt.
Validating the full posteriors produced by photo-z estimators is vital for enabling unbiased cosmology inference with next generation surveys like the LSST \citep{newman2022}.
Synthetic catalogs from PZFlow, together with new metrics of posterior calibration (e.g. the tests of local conditional calibration of \citealt{dey2021b, dey2022}) will be used in future data challenges to optimize and to quantify the error rate and biases of the DESC photo-z estimation pipeline.

In addition to forward modeling, PZFlow is a powerful tool for density estimation applied to tabular data.
We demonstrated this by applying PZFlow to the task of photo-z estimation.
PZFlow achieves high accuracy with very little fine tuning and very few modeling assumptions.
However, as PZFlow is trained via likelihood maximization of the training set, it exhibits mode-covering behavior --- i.e., in order to not miss any modes in the data, PZFlow tends to be conservative and produce overly broad posteriors.
Increasing the amount of training data will likely alleviate these issues, but tradeoffs of this variety are inherent in any choice of machine learning model \citePZp.

While we have developed PZFlow to address the calibration needs of DESC photo-z validation, and have focused on those applications in this paper, we emphasize that PZFlow is a powerful and flexible tool for statistical modeling of any tabular data.

This paper was written using the showyourwork\footnote{\url{https://show-your.work/}} workflow manager.
The code to reproduce this paper is hosted publicly on Github\footnote{\url{https://github.com/jfcrenshaw/pzflow-paper}}, and the code for each individual figure can be found by clicking on the Github logo in the margin next to that figure.

\begin{acknowledgements}
    This paper has undergone internal review in the LSST Dark Energy Science Collaboration.
    We thank the internal reviewers: Martine Lokken, Boris Leistedt, and Benjamin Remy.
    We also thank François Lanusse for some early advice on normalizing flows and to Martine Lokken for testing PZFlow.
    This work was funded by the U.S. Department of Energy, Office of Science, under Award DE-SC0011665.
    J.~F.~Crenshaw, J.~B.~Kalmbach, and A.~J.~Connolly acknowledge support from the DiRAC Institute in the Department of Astronomy at the University of Washington.
    The DIRAC Institute is supported through generous gifts from the Charles and Lisa Simonyi Fund for Arts and Sciences, and the Washington Research Foundation.
    Z.~Yan acknowledges support from the Max Planck Society and the Alexander von Humboldt Foundation in the framework of the Max Planck-Humboldt Research Award endowed by the German Federal Ministry of Education and Research.
    AIM acknowledges support during the course of this work from the Max Planck Society and the Alexander von Humboldt Foundation in the framework of the Max Planck-Humboldt Research Award endowed by the Federal Ministry of Education and Research.

    Author contributions are listed below: \\
    J.~F.~Crenshaw: created PZFlow, designed experiments, wrote paper. \\
    J.~B.~Kalmbach: wrote early normalizing flow code that evolved into PZFlow; photo-z estimation. \\
    A.~Gagliano: validated code, developed use cases and associated tutorials; revised manuscript text. \\
    Z.~Yan: contributed to PZFlow and PhotErr. \\
    A.~J.~Connolly: discussion during development. \\
    A.~I.~Malz: consulted on design and testing of PZFlow. \\
    S.~J.~Schimdt: consulted on design and testing of PZFlow and PhotErr. \\
\end{acknowledgements}

\software{
    adam \citep{adam},
    corner \citep{corner},
    dill \citep{dill},
    jax \citep{jax},
    jupyter \citep{jupyter},
    matplotlib \citep{matplotlib},
    numpy \citep{numpy},
    pandas \citep{pandas,pandas-software},
    scipy \citep{scipy},
    showyourwork \citep{showyourwork},
    scikit-learn \citep{sklearn}
}

\newpage 

\appendix

\section{Training details}
\label{app:training-details}

\begin{figure*}[t!]
    \script{forward_model/plot_galaxy_losses.py}
    \begin{centering}
        \includegraphics{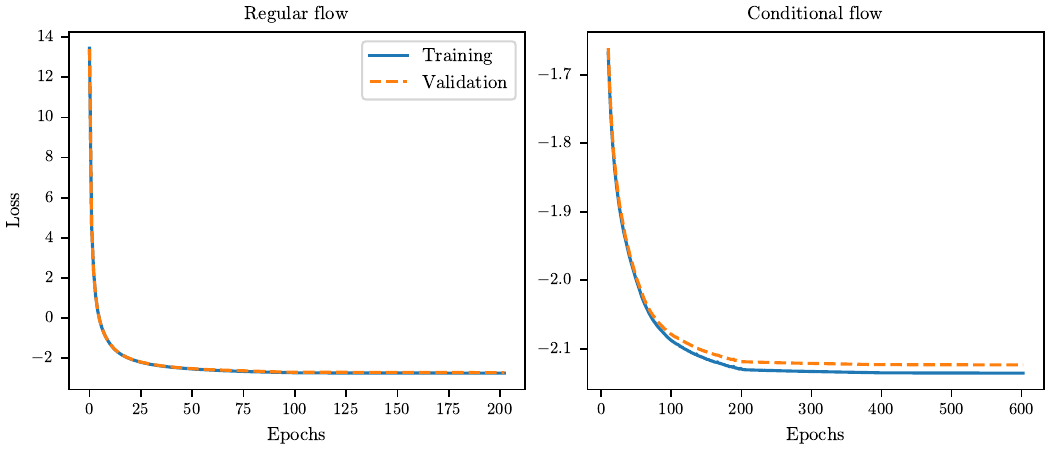}
        \caption{
            Training losses for the normalizing flows used to simulate the galaxy catalog.
            Left: losses for the regular flow.
            Right: losses for the conditional flow.
        }
        \label{fig:galaxy-flow-losses}
    \end{centering}
\end{figure*}

\begin{figure}[t!]
    \script{photo-z/plot_ensemble_losses.py}
    \begin{centering}
        \includegraphics{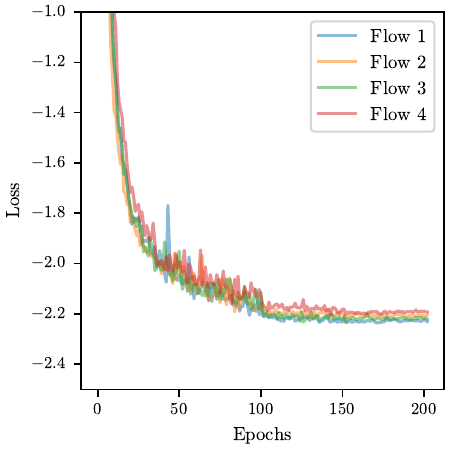}
        \caption{
            Training losses for the four flows in the flow ensemble.
            We have zoomed in to the bottom of the loss curve so you can see that each of the flows converges to a slightly different minimum loss.
        }
        \label{fig:ensemble-losses}
    \end{centering}
\end{figure}

In this section we list some technical details of training the normalizing flows.
Every flow is trained via minimizing the negative log-likelihood
\begin{align}
    \mathcal{L} = - \, \mathbb{E}[ \, \log p(x) \, ],
\end{align}
where the expectation is performed over galaxies in the training set and $p(x)$ is defined in Equation~\ref{eq:px}.

For the main flow in Section~\ref{sec:galaxy-catalog}, we trained for 200 epochs.
We used the Adam optimizer \citep{adam}, starting with a learning rate of $10^{-5}$.
We decreased the learning rate by a factor of 10 after the 100th and 150th epochs.
Training took 7 minutes on a Tesla P100 12GB GPU.
The training loss for this flow is in the left panel of Figure~\ref{fig:galaxy-flow-losses}.

For the conditional flow in~\ref{sec:galaxy-catalog}, we trained for 600 epochs.
Again, we used Adam with an initial learning rate of $10^{-5}$.
We decreased the learning rate by a factor of 10 every 200 epochs.
The training loss for this flow is in the right panel of Figure~\ref{fig:galaxy-flow-losses}.

For each of the flows that make up the flow ensemble in Section~\ref{sec:photo-z}, we trained for 200 epochs using the Adam optimizer.
We started each with a learning rate of $10^{-4}$, which we decreased by a factor of 10 after the 100th and 150th epochs.
The training loss for the ensemble is in Figure~\ref{fig:ensemble-losses}.
Each flow achieved nearly the same training loss.

\section{LSST Error Model}
\label{app:error-model}

We estimate photometric errors for LSST using a generalization of the error model from \citet{ivezic2019}.
To derive the error model, we start with the noise-to-signal ratio (NSR) for an object with photon count $C$ and background noise $N_0$ (which depends on seeing, read-out noise, etc.):
\begin{align}
    \text{NSR}^2 = \frac{N_0^2 + C}{C^2}.
\end{align}
If we define $C=C_5$ when $\text{NSR}= 1/5$, then we can solve for $N_0$ and write
\begin{align}
    \text{NSR}^2 = \frac{1}{C_5} \left( \frac{C_5}{C} \right) + \left[ \left( \frac{1}{5} \right)^2 - \frac{1}{C_5} \right] \left( \frac{C_5}{C} \right)^2.
\end{align}
Defining $x = C_5/C = 10^{(m-m_5)/2.5}$ and $\gamma = 1/5^2 - 1/C_5$, we have
\begin{align}
    \text{NSR}^2 = (0.04 - \gamma) \, x + \gamma \, x^2 ~~ (\text{mag}^2),
\end{align}
which is Equation 5 from \citet{ivezic2019}.
Values for the band-dependent parameter $\gamma$ can be found in Table 2 of the same paper.

In the high signal-to-noise (SNR) limit, $\text{NSR} \ll 1$, and we can approximate
\begin{align}
    \sigma_\text{rand} = 2.5 \log_{10}(1 + \text{NSR}) \approx \text{NSR}.
    \label{eq:err}
\end{align}
This latter approximation is made by \citet{ivezic2019}, and errors are assumed to be Gaussian in magnitude space.
In contrast, we use the exact form of Equation~\ref{eq:err}, and model errors as Gaussian in flux space.
Note that after the photometric errors are applied, the error is re-calculated from the ``observed'' flux, and this new error is reported as the estimated photometric error.
If the original photometric error were reported, it would provide a deterministic link to the original flux.

We have implemented this error model, along with several other extensions, in the Python package PhotErr, which is available on the Python Package Index\footnote{\url{https://pypi.org/project/photerr/}} (PyPI), and Github\footnote{\url{https://github.com/jfcrenshaw/photerr}}.
The extensions include different methods for handling non-detections, methods for modeling errors of extended objects (using models from \citealt{vandenbusch2020,kuijken2019}), and error models for the Roman and Euclid space telescopes \citep{roman,euclid,graham2020}.

\bibliography{bib,nfbib}

\end{document}